\documentclass[pdftex,twocolumn]{aastex62}
\usepackage{amssymb,amsmath}
\usepackage{color}

\graphicspath{{./}{figures/}}

\def\degs{\ifmmode ^{\circ}\else$^{\circ}$\fi}
\def\amin{\ifmmode ^{\prime}\else$^{\prime}$\fi}
\def\asec{\ifmmode ^{\prime\prime}\else$^{\prime\prime}$\fi}
\def\h{$^{\rm h}$}
\def\m{$^{\rm m}$}

\usepackage[normalem]{ulem}  
\usepackage{color,soul}

\usepackage{hyperref}
\usepackage{url}
\usepackage{natbib}

\begin{document}

\title{HUBBLE SPACE TELESCOPE OBSERVATIONS OF THE OLD PULSAR PSR J0108--1431\footnote{Based on observations made with the NASA/ESA {\sl Hubble Space Telescope}, obtained at the Space Telescope Science Institute, which is operated by the Association of Universities for Research in Astronomy, Inc., under NASA contract NAS 5-26555. These observations are associated with program \#14249.}
}

\correspondingauthor{Vadim Abramkin}
\email{vadab94@gmail.com}

\author{Vadim Abramkin}
\author{Yuriy Shibanov}
\affiliation{Ioffe Institute, Politekhnicheskaya 26, St.\ Petersburg, 194021, Russia}
\author{Roberto P.\ Mignani}
\affiliation{INAF -- Istituto di Astrofisica Spaziale e Fisica Cosmica Milano, via E. Bassini 15, I-20133, Milano, Italy}
\affiliation{Janusz Gil Institute of Astronomy, University of Zielona G\'{o}ra, ul Szafrana 2, 65-265, Zielona G\'{o}ra, Poland}
\author{George G.\ Pavlov}
\affiliation{The Pennsylvania State University, Department of Astronomy \& Astrophysics, 525 Davey Lab., University Park, PA 16802, USA}


\begin{abstract}
We present results of optical-UV observations of the 200 Myr old rotation-powered radio 
pulsar J0108$-$1431 with the 
{\sl Hubble Space Telescope}.
We found a putative candidate for the far-UV (FUV) pulsar counterpart, with the flux density $f_\nu = 
9.0\pm 3.2$~nJy at $\lambda = 1528$ \AA.
The pulsar was not detected, however,
at longer wavelengths,
with  $3\sigma$ upper limits of 52, 37, and 87 nJy 
at $\lambda = 4326$, 3355, and 2366 \AA, respectively. 
Assuming that the pulsar counterpart was indeed detected in FUV, and the previously reported marginal $U$ and $B$ detections with the Very Large Telescope were real, 
the optical-UV spectrum of the
pulsar can be described by a power-law model
with a nearly flat $f_\nu$ spectrum. Similar to younger pulsars detected in the optical, the slope of the nonthermal spectrum steepens in the X-ray range. The pulsar's luminosity in the 1500--6000 \AA\ wavelength range, $L \sim 1.2\times 10^{27} (d/210\,{\rm pc})^2$ erg s$^{-1}$, corresponds to a high 
efficiency of conversion of pulsar rotation energy loss rate $\dot {E}$ to the optical-UV radiation, $\eta = L/\dot{E} \sim 
(1$--$6)\times 10^{-4}$, depending on somewhat uncertain values of distance and spectral slope.
The brightness temperature of the bulk neutron star surface does not exceed 59,000 K ($3\sigma$ upper bound), as seen by a distant observer. If we assume that the FUV flux is dominated by a thermal component,
then the surface temperature can be in the range of 27,000--55,000 K, requiring a heating mechanism to operate in old neutron stars.

\end{abstract}

\section{Introduction}
\label{intro}

Optical and ultraviolet (UV) observations of 
old  
rotation-powered 
pulsars (ages $\gtrsim 1$ Myr), 
supplemented by X-ray observations,
are  important to understand advanced stages of  thermal evolution  
of neutron stars (NSs) and study  non-thermal  
emission processes in their magnetospheres. 
So far, only a handful of such pulsars
have been detected in both X-rays and  optical-UV. 
These are 
the 3 Myr old PSR B1929+10 \citep{1996pavlov,2002mlcw} and 
the 17 Myr old PSR B0950+08 \citep{1996pavlov,Pavlov2017}, 
and two a few Gyr old  recycled  millisecond pulsars, PSR J2124$-$3358  \citep{2017rangelov} 
and PSR J0437$-$4715   \citep{2004karg,2012durant},    
all are identified in the UV-optical with the {\sl Hubble Space Telescope} ({\sl HST}). 
For both PSR B1929+10 and PSR J2124$-$3358, the spectral data are insufficient to determine the nature of the optical-UV emission, whereas the others 
show a Rayleigh-Jeans (R-J) continuum, with an additional power-law (PL) component in PSR B0950+08. 
In both cases, the inferred temperatures of $\sim 10^5$ K, higher than predicted by NS cooling models 
\citep[e.g.,][]{2004yp}, 
suggest that some re-heating mechanisms operate in the NS interior. A candidate optical counterpart to the 5 Myr old PSR B1133+16 was found with the Very Large Telescope (VLT), but the identiﬁcation is still uncertain 
\citep{2008zhar,2013zm}.

Another old pulsar with a yet unconfirmed optical counterpart is PSR J0108$-$1431. 
This pulsar was discovered by \citet{1994tauris} 
in the Parkes Southern Pulsar Survey \citep{1996manch}.  
Its spin period $P = 0.808$ s and period derivative
$\dot{P} = 
6.51\times  10^{-17}$ s s$^{-1}$ 
(corrected for the Shklovskii effect), imply a rotational energy loss rate $\dot{E} = 5.1\times 10^{30}$ erg s$^{-1}$ 
and surface magnetic field $B_s = 2.3\times 10^{11}$ G. With the characteristic age of 
196 Myr, PSR J0108$-$1431 is one of the oldest non-recycled isolated radio pulsars known to date. 
It lies close to the so-called ``graveyard'' region in the pulsar $P$-$\dot{P}$ diagram, and it is among the faintest radio pulsars, with a 400 MHz luminosity of 0.391 mJy kpc$^2$ for a distance of $210^{+90}_{-50}$ pc, obtained from the Very Large Baseline Interferometer (VLBI) radio parallax \citep{Deller2009}, 
corrected for the Lutz-Kelker bias \citep{2012verbi}. 

In the first deep optical observation of the field of PSR J0108--1431 with the VLT,
\citet{Mignani2003} 
noticed a faint brightness enhancement in the $U$ image within
 the error ellipse of the Australian Telescope Compact Array (ATCA) radio position, 
projected near the edge of an elongated background galaxy. However, they concluded that most likely it was not a real detection and reported only upper limits in the $V$, $B$ and $U$ filters.

The pulsar position measured by the {\sl Chandra X-ray Observatory} 
by \citet{2009pavlov}
 implied a significant proper motion, which, extrapolated to the epoch of the VLT observations, matched the position of the 
enhancement noticed by  \citet{Mignani2003}. It prompted \citet{Mignani2008}
to propose that the pulsar counterpart had probably been detected with the VLT, with magnitudes $U=26.4\pm 0.3$, $B=27.9\pm 0.5$, $V > 27.8$.
The improved VLBI proper motion,  
$170.0\pm 1.7$ mas yr$^{-1}$ \citep{Deller2009}, made the proposed identification 
more robust, with a chance positional coincidence probability of $\sim 3\times 10^{-4}$  \citep{Mignani2011}.  
The optical spectrum of the PSR J0108--1431 candidate counterpart is poorly defined, although the $U$ and $B$ fluxes are compatible with a $\sim 2.3 \times 10^5$ K R-J spectrum, for the NS radius of 13 km and 210 pc distance. 

The PSR J0108$-$1431 X-ray identification with {\sl Chandra} by 
\citet{2009pavlov} 
has been confirmed by the detection of X-ray pulsations with {\sl XMM-Newton} \citep{Posselt2012}. 
The X-ray spectrum is best fitted by a PL with a (fixed) photon index $\Gamma  = 2$ and a blackbody (BB) with temperature 
1.28$^{+0.35}_{-0.12}\times 10^6$~K and effective radius 
$R= 43^{+16}_{-9} d_{210}$ m, where $d_{210}$ is the pulsar distance in units of 210 pc, with a hydrogen column density $N_H=2.3^{+2.4}_{-2.2}\times 10^{20}$ cm$^{-2}$.

The proposed optical identification of PSR J0108--1431, however, has never been confirmed. Follow-up VLT observations in 2009 with about two times larger total exposures were not conclusive 
because of an almost twice worse seeing  of 0\farcs8 -- 1\farcs0 \citep{Mignani2011},  
as compared to  $\approx$0\farcs5 
in previous  observations in 2000 
 \citep{Mignani2003}.  The counterpart was not detected at the expected new position  of the pulsar accounting for its proper motion, while its  brightness limits,  $U\ga 26.5$, $B\ga 27.2$, were consistent with the tentative detection reported  by \citet{Mignani2008}. 
To verify the putative VLT identification, measure the optical-UV spectrum of this pulsar, and constrain the surface temperature of the very old NS, 
we carried out new observations with the {\sl HST}. We describe the {\sl HST} observations in Section 2 and astrometry of the {\sl HST} images in Section 3.
Photometry of the candidate pulsar counterpart in the {\sl HST} and VLT data is reported in Section 4.
In Section 5 we discuss spectral fits of the optical-UV data, compare them with the X-ray spectrum, and discuss constraints on the NS surface temperature. Conclusions from our analysis are presented in Section 6.

\section{Observations}
\label{obs}
\begin{deluxetable*}{cccrcc}[t]
\tablecaption{{\sl HST} observations of PSR J0108--1431  \label{tab:obs}}
\tablecolumns{6}
\tablenum{1}
\tablewidth{0pt}
\tablehead{
\colhead{Start time\tablenotemark{a}} &
\colhead{Instrument} &
\colhead{Filter} & 
\colhead{$\lambda_p$\tablenotemark{b} } &
\colhead{$\Delta\lambda$\tablenotemark{c} } &
\colhead{Exposure}\\
\colhead{
} & \colhead{} &
\colhead{} &
\colhead{ (\AA)} &\colhead{ (\AA)} &\colhead{(s)} 
}
\startdata 
2016-08-08 15:00:47 & WFC3/UVIS & F225W & 2359 & 500 &2472 \\
2016-08-08 16:33:47 & WFC3/UVIS & F225W & 2359 & 500 &2460 \\
2016-08-08 18:09:13 & WFC3/UVIS & F336W & 3355 & 550 &2580 \\
2016-08-08 19:44:39 & WFC3/UVIS & F336W & 3355 & 550  &2580 \\
2016-08-08 21:20:04 & WFC3/UVIS & F438W & 4325 & 695 &2580 \\
2016-08-08 22:55:30 & WFC3/UVIS & F438W & 4325 & 695 &2580 \\
2016-08-11 22:32:15 & ACS/SBC & F140LP & 1528 & 294 &2800 \\ 
\enddata 
\tablecomments{Each line corresponds to one {\sl HST} orbit.}
\tablenotetext{a}{ UT start time corresponds to the start of first exposure for the orbit.}
\tablenotetext{b}{Pivot wavelength of the filter.}
\tablenotetext{c}{Bandwidth (FWHM) of the filter.}
\end{deluxetable*}
\begin{deluxetable*}{ccccccc}[t]
\tablecaption{{\sl Gaia} DR2 positions and p.m.\ components of the reference objects  
marked in Figure~\ref{fig:1}
\label{tab:ref-stars}}
\tablecolumns{7}
\tablenum{2}
\tablehead{
\colhead{Object\tablenotemark} &
\colhead{RA (J2000)\tablenotemark{a}} &
\colhead{Decl (J2000)\tablenotemark{a}} & \colhead{RA-err} & \colhead{Decl-err} & 
\colhead{$\mu_{\alpha}$\tablenotemark{b}} & \colhead{$\mu_{\delta}$}\\
\colhead{} & \colhead{} & \colhead{} & \colhead{mas} & \colhead{mas} 
& \colhead{mas yr$^{-1}$} & \colhead{mas yr$^{-1}$}
}
\startdata
1 & 01\h08\m09\fs001 & $-$14\degs31\amin32\farcs121 & 0.70 & 0.55 &
+0.6 $\pm$ 1.9 & $-$5.3 $\pm$ 1.2\\
2 & 01\h08\m09\fs181 & $-$14\degs32\amin04\farcs753 & 2.0 & 1.6 &  
\nodata & \nodata \\
3 & 01\h08\m07\fs706 & $-$14\degs32\amin01\farcs488  & 0.15 & 0.12 &  
+5.09 $\pm$ 0.42 & $-$4.12 $\pm$ 0.26\\ 
4 &  01\h08\m15\fs742 & $-$14\degs32\amin33\farcs613 & 0.23 & 0.18 &  
$-$6.56 $\pm$ 0.65 & $-$30.61 $\pm$ 0.41\\
5 & 01\h08\m14\fs793 & $-$14\degs32\amin49\farcs463 & 0.30 & 0.23 &  
+46.72 $\pm$ 0.82 & $-$4.43 $\pm$ 0.52\\
\enddata 
\tablenotetext{a}{The {\sl Gaia} position epoch is 2015.5 (= MJD 57204).} 
\tablenotetext{b}{
$\mu_{\alpha}$ is defined as $\dot{\alpha}\,\cos\delta$, where $\alpha$ and $\delta$ are RA and Decl, respectively.}
\end{deluxetable*}

The PSR J0108--1431 field was observed with the {\sl HST} in 2016 August 
(program \#14249, PI Mignani) using the Ultraviolet-Visible (UVIS) channel of the Wide Field 
Camera 3 (WFC3; 6 {\sl HST} orbits) 
and the Solar Blind Channel (SBC) of the Advanced Camera for Surveys (ACS; 2 {\sl HST} orbits). 
The WFC3/UVIS imaging was carried out with the F438W, F336W and F225W broad-band filters,
while the F140LP long-pass filter was used with the ACS/SBC. 
The log of the observations and the pivot wavelengths of the filters are presented in 
Table~\ref{tab:obs}. 
For each filter, the total integration time  was split 
into shorter exposures distributed over two {\sl HST} 
orbits. 
The WFC3/UVIS exposures were taken in the ACCUM mode, applying a four-point box dither pattern 
in each orbit. 
The UVIS2-C1K1C-CTE aperture was used to place the
pulsar close to a readout amplifier and minimize the CCD charge
transfer efficiency (CTE) losses, as advised in the WFC3
Instrument Handbook\footnote{
See \url{https://hst-docs.stsci.edu/display/WFC3IHB.}}. The data were reduced
and flux-calibrated through the CALWF3 pipeline, which 
also applies the image de-dithering, geometric distortion and CTE corrections,
cosmic-ray filtering, and stacking. For the ACS/SBC,  a single exposure  was taken in the ACCUM 
mode during each orbit. 
The data were processed through the CALACS  pipeline  including the flux-calibration and  
geometric distortion correction. The SBC images are not affected by
cosmic rays and CTE. 
The data for two {\sl HST} orbits obtained for each filter (see Table~\ref{tab:obs}) 
were combined to produce resulting images. However, the aperture door was closed during second ACS/SBC orbit
(2900 s) 
because the Fine Guidance Sensors failed to acquire the guide stars.
Since this orbit provided no science data,
only the first ASC/SBC orbit is included in Table~\ref{tab:obs}.  
  
\section{Astrometry}
\label{astrom}
\begin{figure}[ht]
\begin{center}
\includegraphics[scale=0.263,angle=0]
{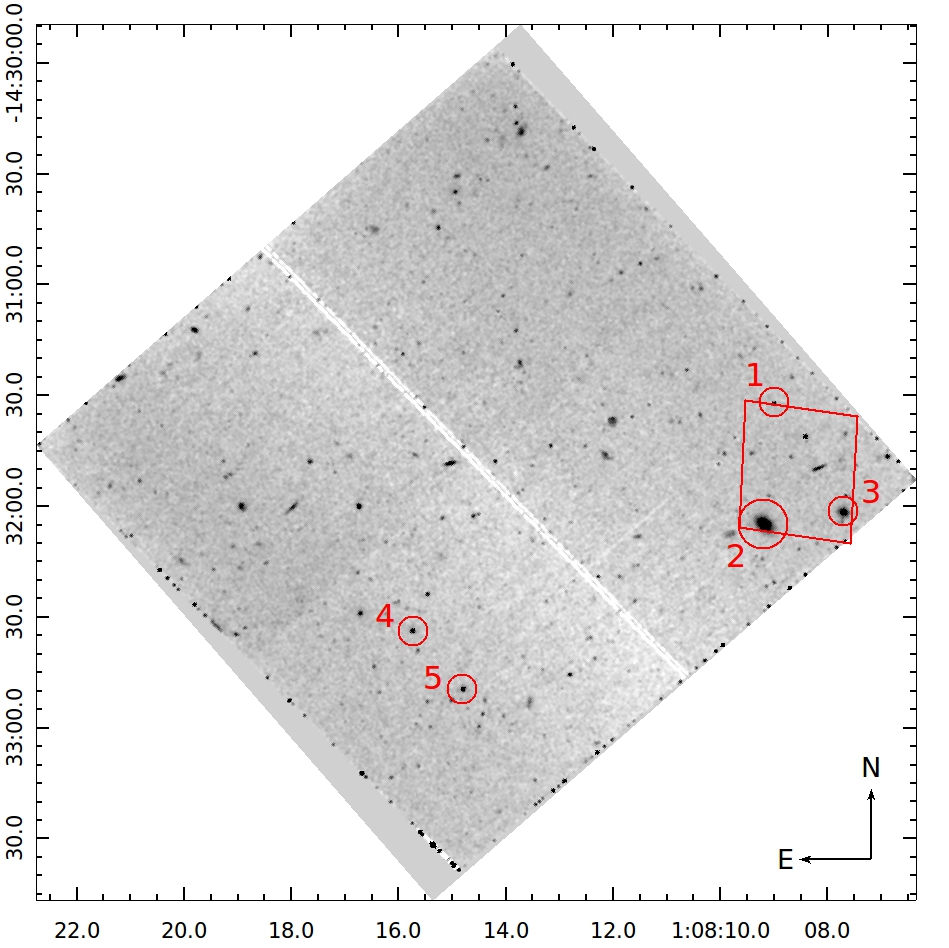} 
\caption{UVIS F438W image of the pulsar field. The image 
is binned by a factor of 8 (new pixel size is 0\farcs32) 
and smoothed with a 2 pixels Gaussian kernel. 
Red numbered circles mark the five  objects from 
the {\sl Gaia} DR2 Catalog listed in Table~\ref{tab:ref-stars} and 
used for astrometry. The red box 
shows the SBC FoV presented in Figure~\ref{fig:2}. 
\label{fig:1}}
\end{center}
\end{figure}

Precise  astrometric referencing is crucial for searching the pulsar counterpart by its positional coincidence 
with the radio pulsar.  
We used the {\sl Gaia} DR2 Catalog \citep{Lindegren2018} and the IRAF tasks {\tt imcentroid} and {\tt ccmap }
to obtain astrometric solutions.   
Five catalogued objects fall within the UVIS field of view (FoV) of about $162\asec \times 162\asec$ 
(pixel scale 
of 39.6 mas was chosen in drizzling). 
These objects are best detected in the F438W image  which we use
as a primary reference image for SBC astrometry
(SBC FoV $\approx 31 \asec \times 34 \asec$; pixel scale 25 mas after drizzling\footnote{For the UVIS and SBC, the original pixel scales are 
about $40\times 40$  and $31\times 34$ mas, respectively.}). 
The {\sl Gaia} objects are marked and numerated 
in this image shown 
in Figure~\ref{fig:1}; 
their coordinates and proper motion (p.m.) components $\mu_{\alpha}$ and $\mu_{\delta}$  
are listed in  
Table~\ref{tab:ref-stars}.
Object 2 looks like an elliptical galaxy and shows no p.m. 

We used the coordinates and p.m.\ values for the four stars in Table~\ref{tab:ref-stars} to calculate their coordinates at the epoch of the UVIS observations, MJD 57608, for further astrometry. To increase the number of reference objects, we also used the galaxy (object 2). Thanks to its regular (elliptical) shape, the uncertainty of its position in the F438W image is reasonably small, about $2.4$ mas.
The total (stars plus galaxy) root-mean-square ({\sl rms}) centroid radial uncertainty is 1.9 mas. According to Table~\ref{tab:ref-stars}, the {\tt rms} of the {\sl Gaia} radial uncertainty of the  reference objects is 1.3 mas. The p.m.\ corrections lead to an additional radial uncertainty of 1.3 mas.
Performing the astrometric fit with the {\tt ccmap}, 
we obtained formal {\tt rms} 
residuals of 0.7 mas for the right ascension (RA) 
and 2.3 mas for the declination (Decl). 
An additional uncertainty of about 2 mas is associated with the 
WFC3/UVIS geometric distortion correction \citep{Kozhurina-Platais2015}. 
As a result, combining all the uncertainties in quadrature,  we  got the final radial uncertainty of 4 mas (0.1 pixels) for astrometric referencing of the F438W image.
\begin{figure}[t!]
\begin{center}
\includegraphics[scale=0.29,angle=0]{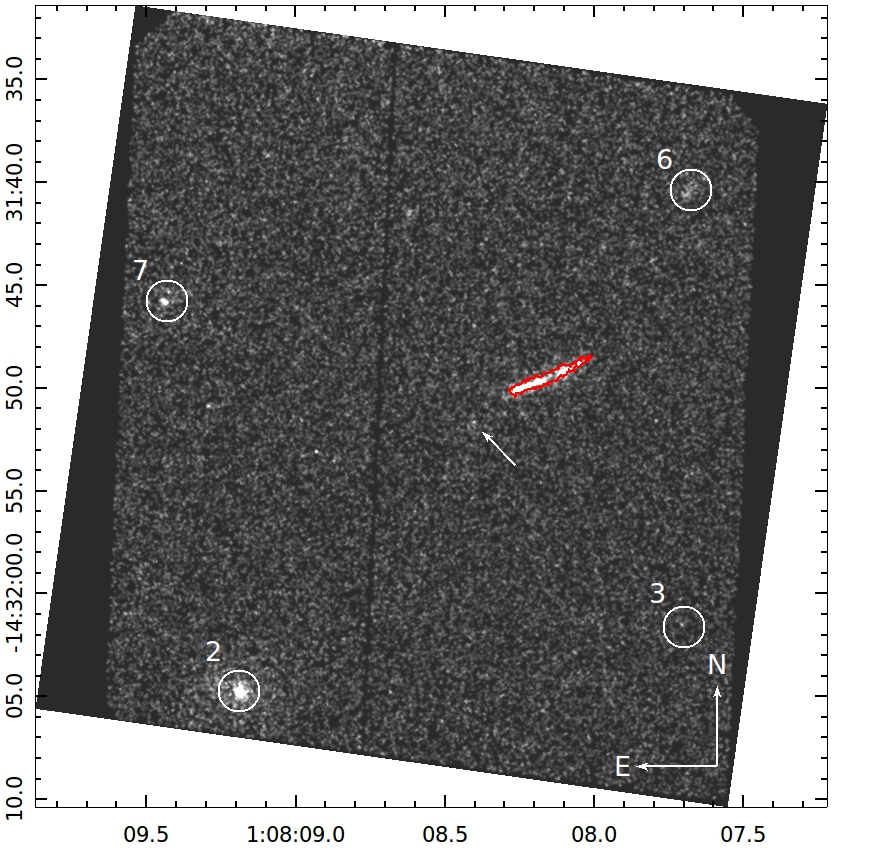}
\caption{SBC F140LP image of the pulsar field. The image is 
binned by a factor of 2 (new pixel size 0\farcs05) and 
smoothed with the 3 pixels Gaussian kernel. 
The arrow shows the pulsar counterpart candidate.
White numbered circles mark four objects  which, together with a bright spiral galaxy seen edge-on north-west  
of the candidate,  were used to align the SBC image to the UVIS F438W image. Red contours  of 
the spiral galaxy 
are overlaid from the F438W image to demonstrate robustness of the alignment.
\label{fig:2}} 
\end{center}
\end{figure} 
     
Using the same reference objects for 
the F336W image yields 
the net radial astrometric uncertainty 
of 7 mas. 
For the F225W image, the {\sl rms} fit 
residuals are 23 mas 
for RA and 13 mas for Decl, 
and the net radial astrometric 
uncertainty is 49 mas. 
The degradation of the referencing accuracy, 
as compared 
to the F438W image, is caused by a significant 
brightness  decrease of the objects   
in the images.

 An accurate alignment of the F140LP and F438W images is not possible with a standard approach because 
three of the only four common objects (marked in Figure~\ref{fig:2}) are extended, with poorly defined positions in the F140LP image. 
The only common point source is star 3, which, however, becomes much fainter in the F140LP band. We used this star 
for an initial estimate of the shifts between the F140LP and F438W images, and found offsets of 1\farcs3  in R.A.\ and 0\farcs4 in Decl.
Their uncertainties  are dominated by the  position error of $\approx$30 mas of star 3 in 
the F140LP image. It was conservatively        
estimated  \citep[see, e.g,][]{1995ApJSNeuschaefer}    
 as FWHM of the SBC point spread function (PSF)   
of $\approx 0\farcs2$ \citep{Avila2016} divided by the signal-to-noise ratio, $S/N\approx 3$, of the star 
in the  image times $\sqrt{8\ln 2} = 2.35$. The {\tt imcentroid} task yields  a similar value.
Then we corrected the WCS values in the header of the SBC fits file 
applying the measured offsets and used the extended objects, particularly the nearby spiral galaxy seen edge-on,
that shows a similar structure in both images, to check the shifts and reveal possible signatures of rotation between the two frames.  
A similar approach was applied by \cite{2002zhar} to align the UV and optical frames for PSR B0950$+$08. Overlaying the frames (see Figure~\ref{fig:2}), we found no signs of additional shifts or
rotation within the  uncertainty of 0\farcs2 
and concluded that the SBC astrometric referencing is confident within this uncertainty.  

\begin{deluxetable*}{ccccc}[t!]
\tablecaption{Positions of PSR 0108$-$1431 at different epochs and 
its p.m.\ components  \label{tab:psr-positions}}
\tablecolumns{5}
\tablenum{3}
\tablewidth{0pt}
\tablehead{
\colhead{Epoch } &\colhead{RA (J2000)} &\colhead{Decl (J2000)} & \colhead{$\mu_{\alpha}$} & \colhead{$\mu_{\delta}$}\\ 
\colhead{MJD} & \colhead{} & \colhead{}  & \colhead{mas yr$^{-1}$} & \colhead{mas yr$^{-1}$}}
\startdata
51752 & 01\h08\m08\fs314(3) & $-$14\degs31\amin49\farcs207(37) & \nodata & \nodata \\
54100 & 01\h08\m08\fs34702(9) & $-$14\degs31\amin50\farcs187(1) & +75.1 $\pm$ 2.3 & $-$152.5 $\pm$ 1.7\\
57608 & 01\h08\m08\fs3967(15) & $-$14\degs31\amin51\farcs651(16) & \nodata & \nodata \\
57611 & 01\h08\m08\fs397(7) & $-$14\degs31\amin51\farcs65(10) & \nodata & \nodata \\
\enddata
\tablecomments{
The coordinates and proper motions of the radio pulsar in the second line are taken from Deller et al.\ (2009). 
They are used to calculate the coordinates at the epochs of the {\sl HST} UVIS and SBC observations (third 
and fourth lines, respectively) and VLT FORS1 observations (first line).
Hereafter, the numbers in brackets are uncertainties related to the last significant digits quoted. 
The uncertainties in the first, third and fourth lines include the pulsar p.m.\ propagation errors 
and the astrometric reference  uncertainties of the corresponding images.} 
\end{deluxetable*}

The most accurate  radio  position and p.m.\ of the pulsar were obtained for the reference epoch of MJD 54100  
with the Very Long Baseline Interferometry  observations  using the Australian Long Baseline Array 
\citep{Deller2009}.  Using them, we calculated the pulsar coordinates 
for the epochs of the {\sl HST} and {\sl VLT} observations 
(Table~\ref{tab:psr-positions}).
The uncertainties on the calculated position in the {\sl HST} images 
(third and fourth lines in 
Table~\ref{tab:psr-positions}) include the uncertainties of the 
F438W and F140LP astrometry   
and the uncertainties 
due to propagation of the pulsar p.m.\ errors.

\section{Possible pulsar counterpart}
\begin{figure}[b!]
\includegraphics[scale=0.318,angle=0]{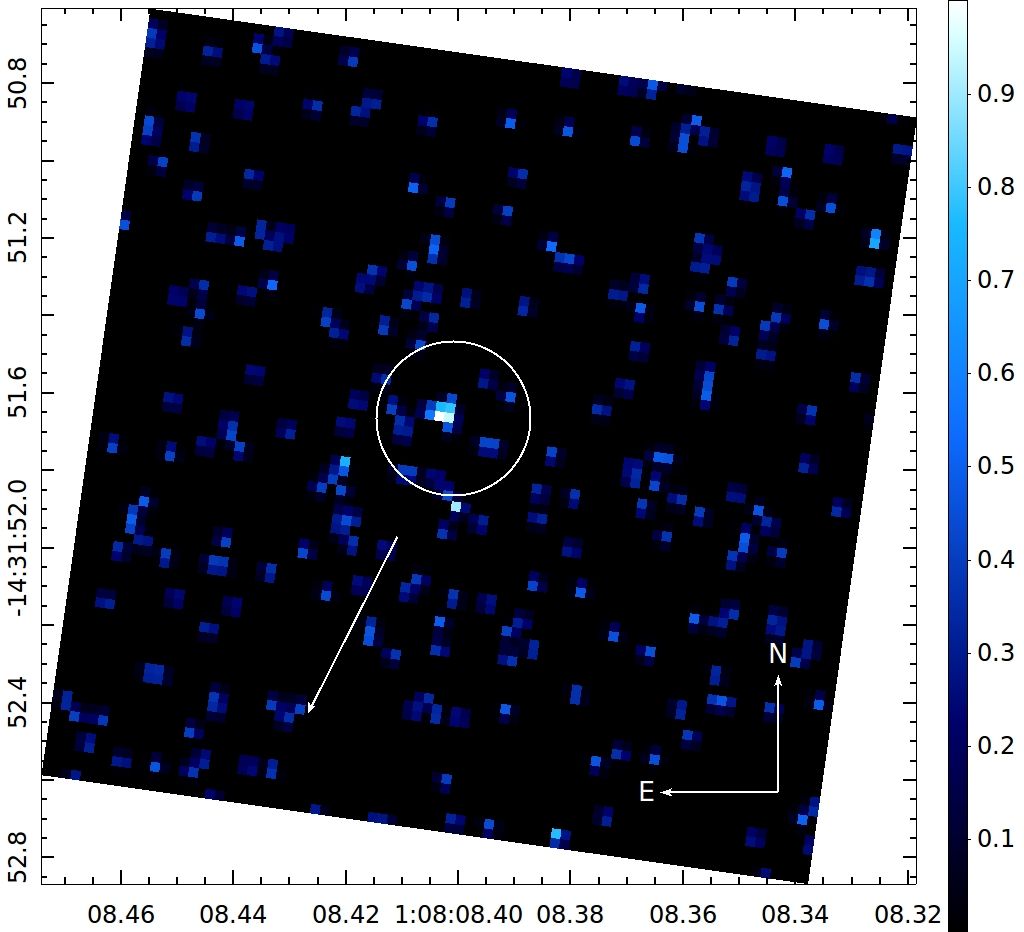}
\caption{2\asec$\times$2\asec region of the 
unbinned and unsmoothed
SBC F140LP image 
around the pulsar. 
The color bar decodes the image intensity in counts per pixel.
The 0\farcs2 radius circle and the arrow mark the 
$1 \sigma$ radio pulsar position uncertainty 
and the direction of its p.m. A faint point-like 
counterpart candidate to the pulsar is clearly visible within the circle.
\label{fig:4}}
\end{figure}

In the SBC/F140LP image (Figure~\ref{fig:2})   
we found a faint point-like source with coordinates  
RA = 01\h08\m08\fs403(14) and  
Decl = $-$14\degs31\amin51\farcs66(20),    
 consistent with the expected  radio pulsar  
coordinates (see 4th row in Table~\ref{tab:psr-positions}).
A zoomed-in region around this source is shown in  Figure~\ref{fig:4}, which 
clearly demonstrates  that  the source is located 
within the circle  with the radius 
of 0\farcs2 corresponding to the 1$\sigma$ uncertainty of the pulsar radio position in this image. 


\begin{deluxetable*}{ccccccc}[ht]
\tablecaption{SBC photometry of the pulsar counterpart candidate \label{tab:photometry-mod}}
\tablenum{4}
\tablehead{
\colhead{$N_{t}$} & \colhead{$C_{\rm pos}$} 
& \colhead{$\overline{C}_{\rm bgd}$} & \colhead{$\sigma_{\rm bgd}$}& \colhead{$N_{s}$} & \colhead{$C_{s}$} & 
\colhead{$f_{\nu}$}\\
\colhead{(cts)} & \colhead{(cts/ks)} 
& (cts/ks) & \colhead{(cts/ks)} & 
\colhead{(cts)} & \colhead{(cts/ks)} & \colhead{(nJy)}
}
\startdata
 15.3 & 5.5 
 & 2.0 & 0.7 & $9.8\pm 3.7$ & $5.6\pm 2.0$ & $9.0\pm 3.2$
\enddata
\tablecomments{
$N_{t}$ is the total number of counts in the $A_s=0.126$ arcsec$^2$ source aperture, $C_{\rm pos}= N_t/t_{\rm exp}$ is the 
count rate measured in the source aperture ($t_{\rm exp}=2.8$ ks is the exposure time), 
$\overline{C}_{\rm bgd}$ and $\sigma_{\rm bgd}$ are the mean and standard deviation of background measurements in 1000 circles with the 0\farcs2 radius in the $A_b=4.12$ arcsec$^2$ annulus,   $N_{s}$
is the net source count number in the source aperture,
its error is estimated as  
$[(\sigma_{\rm bgd} t_{\rm exp})^2 + N_s]^{1/2}$,  
$C_{s}=N_s/t_{\rm exp}/\phi_E$ is the aperture-corrected net 
source count rate, $\phi_E\approx 0.63$ \citep{Avila2016} is a fraction of source counts in the $r=0\farcs2$ aperture, $f_\nu = C_s {\cal P}_\nu$ is the flux density at the pivot wavelength $\lambda_{\rm piv}=1528$ \AA, and
${\cal P}_\nu = 1.61$ nJy ks cts$^{-1}$ 
 is 
the count-rate-to-flux conversion factor.
}
\end{deluxetable*}

\subsection{Photometry of the FUV counterpart candidate}

To find an optimal source aperture for the SBC/F140LP photometry of the pulsar counterpart candidate,
we calculated the signal-to-noise ratio 
${\rm S/N}$ as a function of  radius of 
circular aperture centered at the brightest source pixel in the unbinned image using 
the background extracted from the annulus around the same center with inner and outer radii of 20 
and 50 
pixels 
(area $A_b=4.12$ 
arcsec$^2$). 
We found the maximum S/N = 2.7 in the aperture 
of 8 pixels (0\farcs2) radius
(area $A_s=0.126$ arcsec$^2$). This aperture was selected as the optimal one. 
According to Table 2 in \citet{Avila2016}, 
it contains the fraction $\phi_E\approx 0.63$ of the total number of point source counts. 
Following \citet{Guillot2019}, we estimated the mean background $\overline{C}_{\rm bgd}$ and its standard deviation $\sigma_{\rm bgd}$
for a set of 1000 circular background  
regions of size $r_{\rm extr}$=0\farcs2 (8 pixels, the same as of the source aperture), randomly 
placed in the annulus. 
To convert the count rate to the flux density, we took into account the recent correction of the SBC sensitivity, such that 
the flux density at a given count rate is about 0.77 of the previously adopted value \citep{Avila2019}.
The results are presented in  
Table \ref{tab:photometry-mod}. 



As the putative pulsar counterpart 
is detected at only about $3\sigma$ level, we cannot rule out that the count 
rate excess at the pulsar position is caused by some fluctuation.
In this case, one can estimate the flux density upper bound following \citet{Kashyap2010} and \citet{Guillot2019}.
We define the upper bounds $C_{\rm ub}$ on the pulsar count rate as $C_{\rm ub}=C_{\rm pos}-\overline{C}_{\rm bgd}+n\sigma_{\rm bgd}$,
where 
$C_{\rm pos}$ is the measured count rate at the position of the pulsar, $\overline{C}_{\rm bgd}$ and $\sigma_{\rm bkg}$ are defined above (see also 
Table \ref{tab:photometry-mod}),
and $n$ determines the significance level of the upper bound\footnote{This 
definition of $C_{\rm ub}$ is applicable at $C_{\rm pos} \geq \overline{C}_{\rm bgd}$, which is fulfilled 
in our case.}.
The resulting 3$\sigma$ upper bounds 
are 
$C_{\rm ub}=5.5$ cts/ks and  
$f_\nu^{\rm ub} = 14$ nJy.
\begin{deluxetable*}{cccccccccccc}[t]
\tablecaption{
Upper bounds on count rates and mean flux densities
in the UVIS filters at the pulsar radio position, $\alpha$=01\h08\m08\fs397 and $\delta$=$-$14\degs31\amin51\farcs65, at the {\sl HST} observations epoch (MJD 57608)\label{tab:upperbounds}}
\tablecolumns{11}
\tablenum{5}
\tablehead{
\colhead{Filter} &
\colhead{$\lambda_{\rm piv}$} &
\colhead{$t_{\rm exp}$} &
\colhead{$r_{\rm extr}$} & \colhead{$\phi_{E}$} & \colhead{$C_{\rm pos}$} & \colhead{$\overline{C}_{\rm bgd} \pm \sigma_{\rm bgd}$} & \colhead{$C_{\rm ub}$} & \colhead{${\cal P}_{\nu}$} & \colhead{$f_{\nu,3\sigma}^{\rm ub}$} & \colhead{$f_{\nu,1\sigma}^{\rm ub}$}\\
\colhead{} & \colhead{(\AA)} & \colhead{(s)} & \colhead{($''$)} & \colhead{(\%)} & \colhead{(cts/ks)} & \colhead{(cts/ks)} & \colhead{(cts/ks)} & \colhead{(nJy ks/cts)} & \colhead{(nJy)} & \colhead{(nJy)}
}
\startdata
F438W & 4326 & 5160 & 0.14 & 81 & 47 & 
$16\pm 24$ & 
102 & 0.416 & 52 & 28\\
F336W & 3355 & 5160 & 0.12 & 78 & 
27 & 
$13\pm 16$ & 
61 & 0.470 & 37 & 20\\
F225W & 2366 & 4932 & 0.14 & 74 & 
46 & 
$16\pm 17$ & 
82 & 0.783 & 87 & 50\\
\enddata
\tablecomments{
 $C_{\rm ub}$ and $f_{\nu,3\sigma}^{\rm ub}$ are the $3 \sigma$ 
 upper bounds; $f_{\nu,1\sigma}^{\rm ub}$ is the $1 \sigma$ upper bound. 
 }
 
\end{deluxetable*}
\begin{deluxetable*}{cccccccccc}[t!]
\tablecaption{$3\sigma$ upper bounds on count rate and mean flux density 
at the pulsar radio position, $\alpha$=01\h08\m08\fs314 and $\delta$=$-$14\degs31\amin49\farcs207, at the VLT observations epoch (MJD 51752)
\label{tab:upperbounds2}}
\tablecolumns{10}
\tablenum{6}
\tablehead{
\colhead{Filter} &
\colhead{$\lambda$} &
\colhead{$t_{\rm exp}$} &
\colhead{$r_{\rm extr}$} & \colhead{$\phi_{E}$} & \colhead{$C_{\rm pos}$} & \colhead{$\overline{C}_{\rm bgd} \pm \sigma_{\rm bgd}$} & \colhead{$C_{\rm ub}$} & \colhead{${\cal P}_{\nu}$} & \colhead{$f_{\nu,3\sigma}^{\rm ub}$}\\
\colhead{} & \colhead{(\AA)} & \colhead{(s)} & \colhead{($''$)} & \colhead{(\%)} & \colhead{(cts/ks)} & \colhead{(cts/ks)}  & \colhead{(cts/ks)} & \colhead{(nJy ks/cts)} & \colhead{(nJy)}
}
\startdata
F438W & 4326 & 5160 & 0.14 & 81 & 
13 & 
$15\pm 17$ & 
50 & 0.416 & 26\\
F336W & 3355 & 5160 & 0.12 & 78 & 6.4 & 
$12 \pm 13$ & 
40 & 0.470 & 24\\
F225W & 2366 & 4932 & 0.14 & 74 & 
25 & 
$6\pm 13$ & 
59 & 0.783 & 62\\
F140LP & 1528 & 2800 & 0.2 & 63 & 0.9 & 2.1 $\pm$ 0.6 & 1.8 & 1.61 & 5\\
\enddata
\tablecomments{For the F438W, F336W and F140LP bands, $C_{\rm pos}<\overline{C}_{\rm bgd}$ 
and the upper bounds 
are calculated as $C_{\rm ub}=3\sigma_{\rm bgd}$.
} 
\end{deluxetable*}
\subsection{UVIS upper bounds and VLT observations of PSR J0108-1431 \label{UVIS+VLT}}

The counterpart candidate is not detected in the UVIS images.
Therefore, we calculated the upper bounds on the pulsar count rate and flux density  for each of the UVIS filters  using the same approach as for the F140LP filter.
The sizes of the extraction regions used to measure $C_{\rm pos}$ were chosen by identifying in each image 
the extraction radius that maximizes  $\rm S/N$ 
for a point source (star). 
For each of the filters, we estimated $\overline{C}_{\rm bgd}$ and $\sigma_{\rm bgd}$ from a set of 1000 circular background  
regions of size $r_{\rm extr}$, randomly selected 
in the 
annulus of 10 pixels 
and 30 pixels inner and outer radii 
around the pulsar position. 
These quantities, together with $C_{\rm ub}$ and the corresponding flux densities corrected for the finite extraction aperture, $f_\nu^{\rm ub} = C_{\rm ub} {\cal P}_\nu \phi_E^{-1}$, are presented in Table~\ref{tab:upperbounds} where we also show the $1\sigma$ flux density  upper bounds used in spectral fits (Section 5.2).

We also do not detect any object in any of the {\sl HST} bands  at the
position where \citet{Mignani2008} found a possible pulsar counterpart in the VLT $U$ and $B$  bands near the northern edge of the spiral galaxy. 
\begin{figure*}[t]
\begin{minipage}[h]{0.5\linewidth}
\includegraphics[scale=0.3,angle=0]{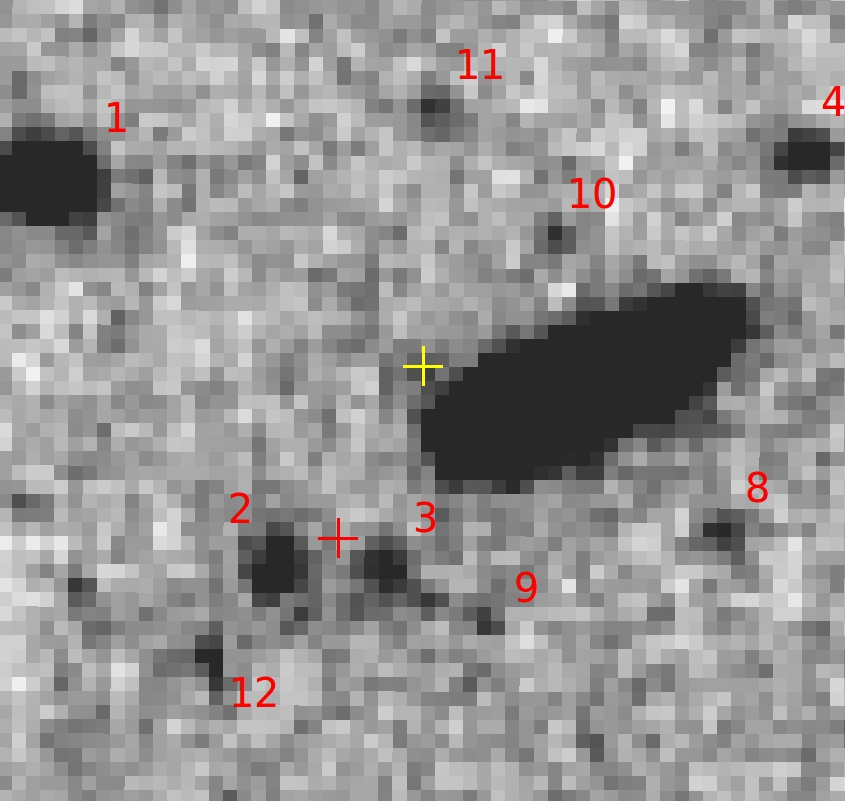}
\end{minipage}
\begin{minipage}[h]{0.5\linewidth}
\includegraphics[scale=0.3,angle=0]{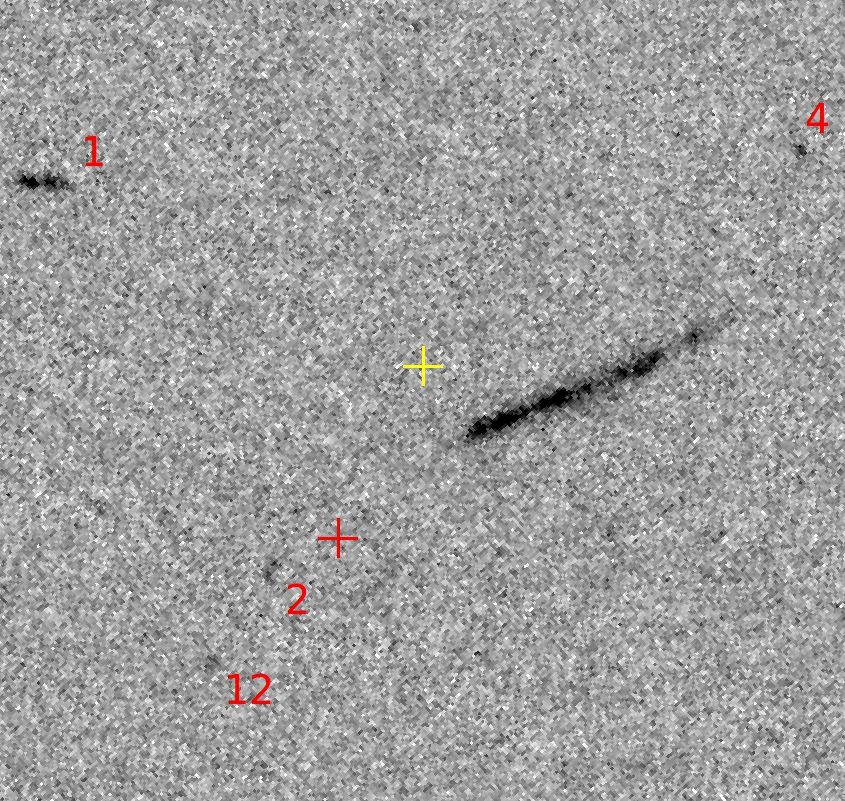}
\end{minipage}
\caption{
12\farcs0$\times$11\farcs5\ vicinity of the pulsar as seen with the VLT in the $U$$+$$B$ band ({\sl left:}) and  with the {\sl HST}/UVIS in the $F438W$$+$$F336W$ band 
({\sl right:}). North is up and east is left. Smoothing with the Gaussian kernel 
radius of 1 pixel is applied to the VLT image.  The sources are numerated using the nomenclature of \citet{Mignani2003}.   
Pulsar radio positions at the VLT epoch (2000) and the {\sl HST}  epoch (2016) are shown by  yellow and red crosses, respectively. 
A faint source  is marginally detected in the VLT image near the 2000 position of the pulsar, while it is absent in the {\sl HST} image (for details see Sect.~\ref{UVIS+VLT})  
\label{fig:6}}
\end{figure*}
For the {\sl HST}--VLT consistency check, we re-reduced  the VLT $UBV$  data 
using  the recent version of the  
ESO recipe execution tool EsoRex 3.13.12\footnote{\url{https://www.eso.org/sci/software/cpl/esorex.html}}. To maximize the spatial resolution and better resolve the putative counterpart from the  galaxy, we selected six best-seeing 900 s  exposures of eight available in $B$, three 1800 s exposures of five available in $U$, and all twelve  
600 s exposures 
in $V$\footnote{For the VLT observation log, see \citet{Mignani2003}.}. This  resulted in an extremely good seeing on the stacked images of 0\farcs56 in the $U$ and $B$ bands and 0\farcs54 in the $V$ band. 

The VLT astrometric solution was also revised using 11 relatively bright unsaturated {\sl Gaia} stars, 
with account for their p.m.\ shifts for  the epoch of the VLT observations.  This   
resulted in a 30 mas uncertainty  of the WCS  referencing of all images  in both coordinates, which is significantly better than the previous  astrometric 
precision of 190 mas 
based on the GSC-II catalog \citep{Mignani2008}. 
The more precise astrometry  
supports the assumption that the object detected with the VLT 
at a $2\sigma$ significance  in the $U$ and $B$ bands is the counterpart candidate.
Stacking the $U$ and $B$ images 
increases the detection significance to $3\sigma$.
The region of this image containing the pulsar  is shown in the left panel of Figure~\ref{fig:6}. 
For comparison, we also merged the {\sl HST}/UVIS images in the F336W and F438W bands.  
The respective region   
is  presented in the right panel of Figure~\ref{fig:6}. 
Not only the VLT source proposed  as the pulsar counterpart is not seen in the F336W+F438W image but also some other faint 
VLT objects, such as sources 3 and 8--11  
in \citet{Mignani2003},  are also either resolved only marginally or not visible.
At the same time, bright structures of the spiral galaxy are better resolved with the {\sl HST}.  

Photometric calibration of the VLT data was carried out using Landolt's standards PG1323--086, PG0231+051 and PG2331+055, observed at the same nights as the target; it resulted in the following  magnitude zero-points for the stacked images: $Z_U=25.08\pm0.04$,  $Z_B=27.55\pm0.05$ and $Z_V=28.00\pm0.01$,  
%
calculated for the fluxes in units of the CCD electron rate.  
For photometry of the putative pulsar counterpart, we used apertures of 2.5 pixel in $U$ and 1.3 pixel  in $B$ (the pixel scale was 0\farcs2) which correspond to the enclosed energy fraction of 0.67 and 0.3, respectively, measured using 
bright unsaturated stars.
Such small apertures were used 
because of proximity of the bright spiral galaxy leading  to large systematic errors. We obtained  flux densities of the source of $44\pm22$ nJy   and $30\pm14$ nJy in the $U$ and $B$ bands, respectively, and a $3\sigma$ upper bound of 36 nJy in the $V$ band.  
The fluxes are consistent, within the uncertainties,
with the results obtained by \cite{Mignani2008}. 

To understand the nature of the putative VLT pulsar 
counterpart,  
we calculated the {\sl HST} upper bounds at its position in the VLT observations epoch
(Table~\ref{tab:upperbounds2}). 
The derived flux density 
$3\sigma$ upper bounds of 24 nJy in the F336W band and 26 
nJy in the F438W band 
are somewhat lower than the flux densities  in the VLT $U$ and $B$ bands, respectively, presented above.
This  implies that
the VLT object  has disappeared or moved out of its place 
by the epoch of the {\sl HST} observations, i.e., it is not a steady field object. 
On the other hand, our $3\sigma$ upper bounds of 37 nJy in the F336W band  and 52 nJy in the F438W band 
at the pulsar position at the {\sl HST} epoch (Table~\ref{tab:upperbounds}) are 
comparable to the $U$ and $B$ band flux densities\footnote{ Difference of the upper bounds 
at the two pulsar positions is due to different properties of the local backgrounds.}. 
This suggests that the faint pulsar counterpart  
was indeed seen in  the $U$, $B$ and F140LP bands, 
and it 
could be seen in the {\sl HST} F336W and F438W bands 
if the exposures were 
just a factor of 1.5 longer. 


Thus, we cannot rule out the possibility that both the VLT ($U$ and $B$ bands) and {\sl HST} (F140LP band) detected the pulsar counterpart, but further deep observations are needed to prove it.

\section{Discussion \label{disc}}
We likely detected the putative pulsar FUV counterpart in 
the F140LP band. Our photometric measurements
show that its brightness  is near the SBC detection threshold 
for the one-orbit {\sl HST} 
observation.
Three {\sl HST} orbits would be 
needed to confirm the FUV counterpart at a 
$5\sigma$  significance. 
Nevertheless, the tentative detections with the {\sl HST} ACS/SBC and VLT FORS1 $U,B$ bands, combined 
with the 
upper bounds obtained in the UVIS 
(and VLT $V$) bands,   
as well as the {\sl Chandra} and {\sl XMM-Newton} X-ray  data,  can provide interesting constraints on the optical-UV-X-ray spectral energy distribution (SED) of this old  pulsar.      

\begin{deluxetable*}{ccccccccccc}[t]
\tablecaption{
De-reddened flux densities  
and $3\sigma$($1\sigma$)
flux density  bounds of the pulsar counterpart (in nJy), and PL fit parameters,
for three values of 
$E(B-V)$.
\label{tab:extinction}}
\tablenum{7}
\tablehead{
 \colhead{$E(B-V)$} &
\colhead{$f_U$} &
\colhead{$f_B$} &
\colhead{$f_V$} & \colhead{$f_{\rm F140LP}$} & \colhead{$f_{\rm F225W}$} & \colhead{$f_{\rm F336W}$} & \colhead{$f_{\rm F438W}$} & \colhead{$\alpha$} & \colhead{$f_0$} }
\startdata
0.01 & $46 \pm 23$ & $31 \pm 14$ & $<37 (26)$ 
& $9.7\pm3.4$ & $<94 (54)$ & $<39 (21)$ & $< 54 (29)$ & $-0.71^{+0.74}_{-0.41}$ & $16.2^{+3.3}_{-7.1}$ \\
0.02 & $48 \pm 24$ & $33 \pm 16$ & $<38 (27)$
& $10.5 \pm 3.7$ & $<101 (58)$ & $<41 (22)$ & $<56 (30)$ & $-0.63^{+0.81}_{-0.45}$ & $16.7^{+3.5}_{-7.8}$ \\
0.03 & $50 \pm 25$ & $34 \pm 16$ & $<39 (27)$
& $11.3 \pm 4.0$ & $<109 (63)$ & $<43 (23)$ & $<58 (31)$ & $-0.60^{+0.77}_{-0.44}$ & $17.7^{+3.6}_{-8.2}$ \\
\enddata
\tablecomments{
Extinction values $A_\lambda$, used for de-reddening,  were calculated using using the extinction law from  \citet{Cardelli1989}.
HST upper bounds were calculated at the pulsar radio position, $\alpha$=01\h08\m08\fs397 and $\delta$=$-$14\degs31\amin51\farcs65, at the {\sl HST} observations epoch (MJD 57608).
}
\end{deluxetable*}

\subsection{Extinction towards PSR J0108--1431}
According to the Galactic 3D extinction map by \citet{Green2018},
the color excess $E(B-V)$ varies between 0.00 and 0.04 within the uncertainty of the distance to the pulsar, 
and $E(B-V)=0.02^{+0.02}_{-0.01}$ 
for  $d=210$ pc.  

The $E(B-V)$ value is correlated  
with the 
effective hydrogen column density $N_{\rm H}$
for X-ray photoelectric absorption models. 
Applying the empirical relation $N_{\rm H}=(0.7\pm0.1) \times 10^{22} E(B-V)$~cm$^{-2}$, obtained by \citet{Watson2011} for the Galaxy using observations of X-ray afterglows of a large number of $\gamma$-ray bursts, we 
expect $N_{\rm H}=1.4^{+1.8}_{-0.8}\times 10^{20}$~cm$^{-2}$.

Alternatively, $N_{\rm H}$
can be estimated 
using the pulsar's dispersion measure, ${\rm DM} = 2.38$ pc\,cm$^{-3}$ and the correlation between  ${\rm DM}$ and $N_{\rm H}$ obtained by \citet{He2013}, 
 $N_{\rm H}= 0.30^{+0.13}_{-0.09} \times 10^{20} {\rm DM}$~cm$^{-2}$, which yields 
$N_{\rm H}=0.71^{+0.31}_{-0.21}\times 10^{20}$~cm$^{-2}$, in 
agreement with the value derived from $E(B-V)$. 
This also overlaps with the $N_{\rm H}$ range of $(0.3-0.8)\times 10^{20}$~cm$^{-2}$ at $d=210$ pc obtained from 
the study of interstellar NaD absorption lines \citep{posselt2007,posselt2008}. 

Finally, the 
phase-integrated X-ray spectrum 
of the pulsar obtained with \textit{XMM-Newton} is 
most plausibly described by the absorbed power 
law (PL) plus blackbody (BB) model with $N_{\rm H} = 2.3^{+2.4}_{-2.3}\times 10^{20}$~cm$^{-2}$ \citep{Posselt2012,Arumugasamy2019}. The latter value is very uncertain but  consistent with the above estimates. 
\begin{figure*}[t!]
\begin{minipage}[h]{0.6\linewidth}
\includegraphics[scale=0.23,angle=0]{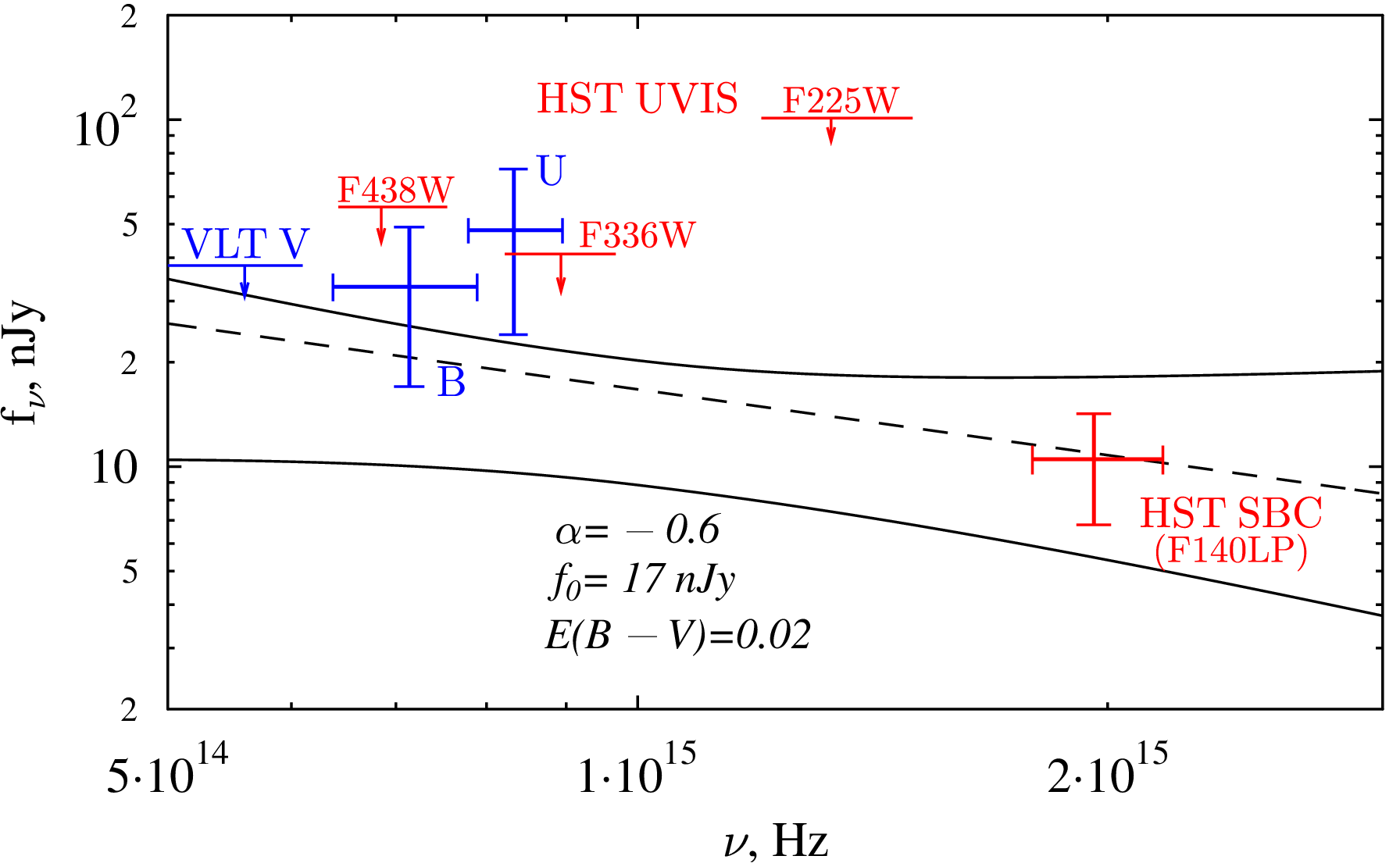}
\end{minipage}
\hfill
\begin{minipage}[h]{0.4\linewidth}
\includegraphics[scale=0.25,angle=0]{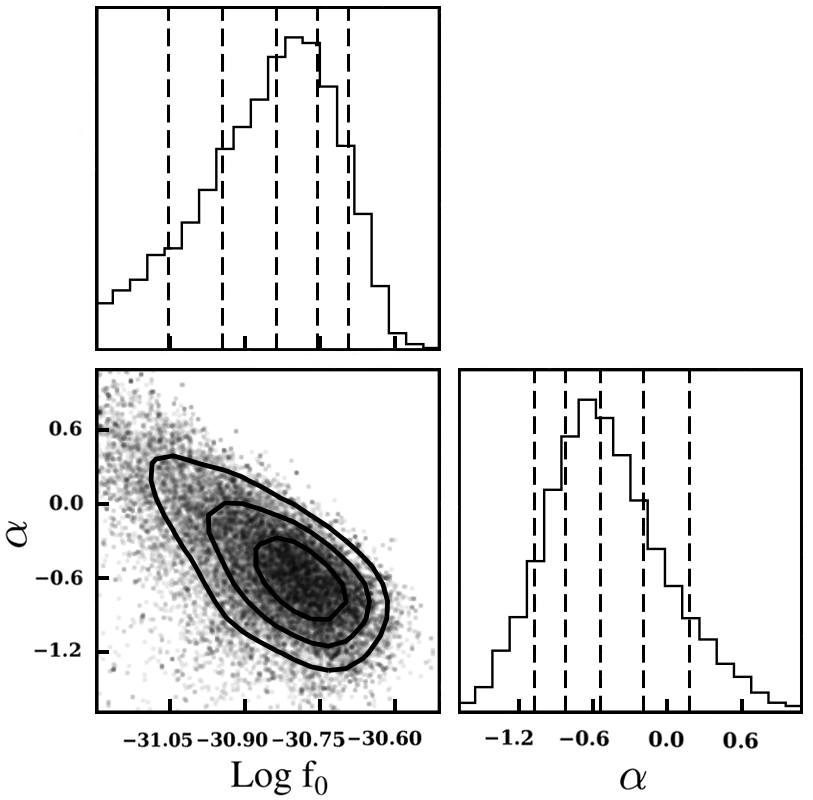}
\end{minipage}
\caption{
{\sl Left:} 
Unabsorbed optical-UV SED  
of the pulsar counterpart candidate for $E(B-V)=0.02$. 
Red  and blue  error bars and downward arrows show  flux densities and $3\sigma$
upper bounds 
measured with the {\sl HST} and VLT, respectively, with instruments and 
filters  indicated in the plot.
The dashed and solid lines correspond to the best fit of the data by the PL model and its 90\% credible  uncertainty, respectively.  
{\sl Right:} 2D and 1D marginal posterior 
probability distributions for the PL model parameters $\log f_0$ and $\alpha$ ($f_0$ in units of erg\,cm$^{-2}$\,s$^{-1}$\,Hz$^{-1}$).  
Vertical dashed lines in the 1D plots correspond 
to the 10\%, 25\%, 50\%, 75\%, and 90\% percentiles of the distributions. The contours in the 2D plot 
represent the  levels at 75\%, 50\%, and 25\% of the maximum probability value.
\label{fig:7}}
\end{figure*}

All in all, we can accept  
$E(B-V)=0.01$--0.03 as the most probable color excess range for 
de-reddening of the optical-UV data and combining them consistently with the X-ray data. 
Using this color excess range and the extinction 
law from \citet{Cardelli1989},
 we can calculate the extinction 
$A_{\lambda}$, and the  de-reddened source flux density  
or its upper bound for all the bands where the pulsar was observed. 
These quantities are  presented in Table~\ref{tab:extinction} and Figure~\ref{fig:7}.

\subsection{Multi-wavelength  spectrum of PSR J0108--1431}
Optical-UV emission from a rotation powered pulsar generally consists of two components, thermal and nonthermal. The nonthermal component, produced by relativistic particles in the pulsar magnetosphere, is usually described by a power-law (PL) model, $f_\nu \propto \nu^\alpha$, while the spectrum of the thermal component, emitted from the 
NS surface, is close to a blackbody spectrum \citep{2011mignani}. In middle-aged 
(0.1--1~Myr) and moderately 
old (1--10~Myr) pulsars the nonthermal component dominates in the optical 
while the thermal component dominates in FUV 
\citep[e.g.,][]{2001kopts,2004zhar,2006shib,Kargaltsev2007, Pavlov2017},
However, we know too little about optical-UV emissions of pulsars as old as PSR J0108--1431; the only other very old pulsar, J2144--3933 with the age of 300 Myr,
observed in the optical-UV, was not detected \citep{Guillot2019}. Therefore, we should explore various options and their connection with the results 
in X-rays where spectra of very old pulsars, including PSR J0108--1431,  
typically show only the PL component of the NS magnetosphere origin and a thermal component 
from small hot  spots  at the NS surface near magnetic pole regions  heated by relativistic particles generated 
in its magnetosphere. 

\subsubsection{Possible power-law spectrum of the tentative pulsar counterpart}
\begin{figure*}[tbh!]
\begin{minipage}[h]{0.5\linewidth}
\includegraphics[scale=0.2,angle=0]{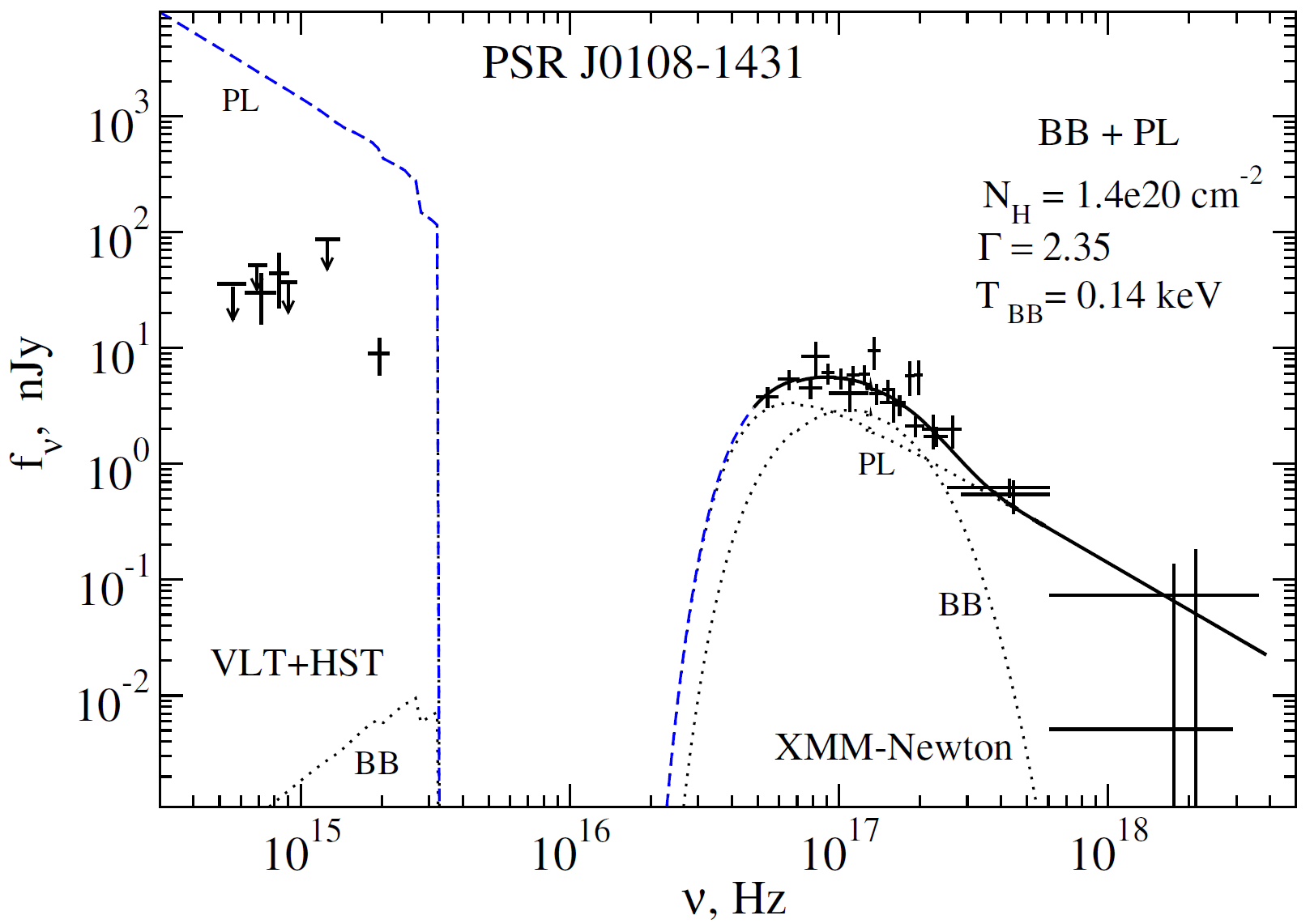}
\end{minipage}
\hfill
\begin{minipage}[h]{0.5\linewidth}
\includegraphics[scale=0.195,angle=0]{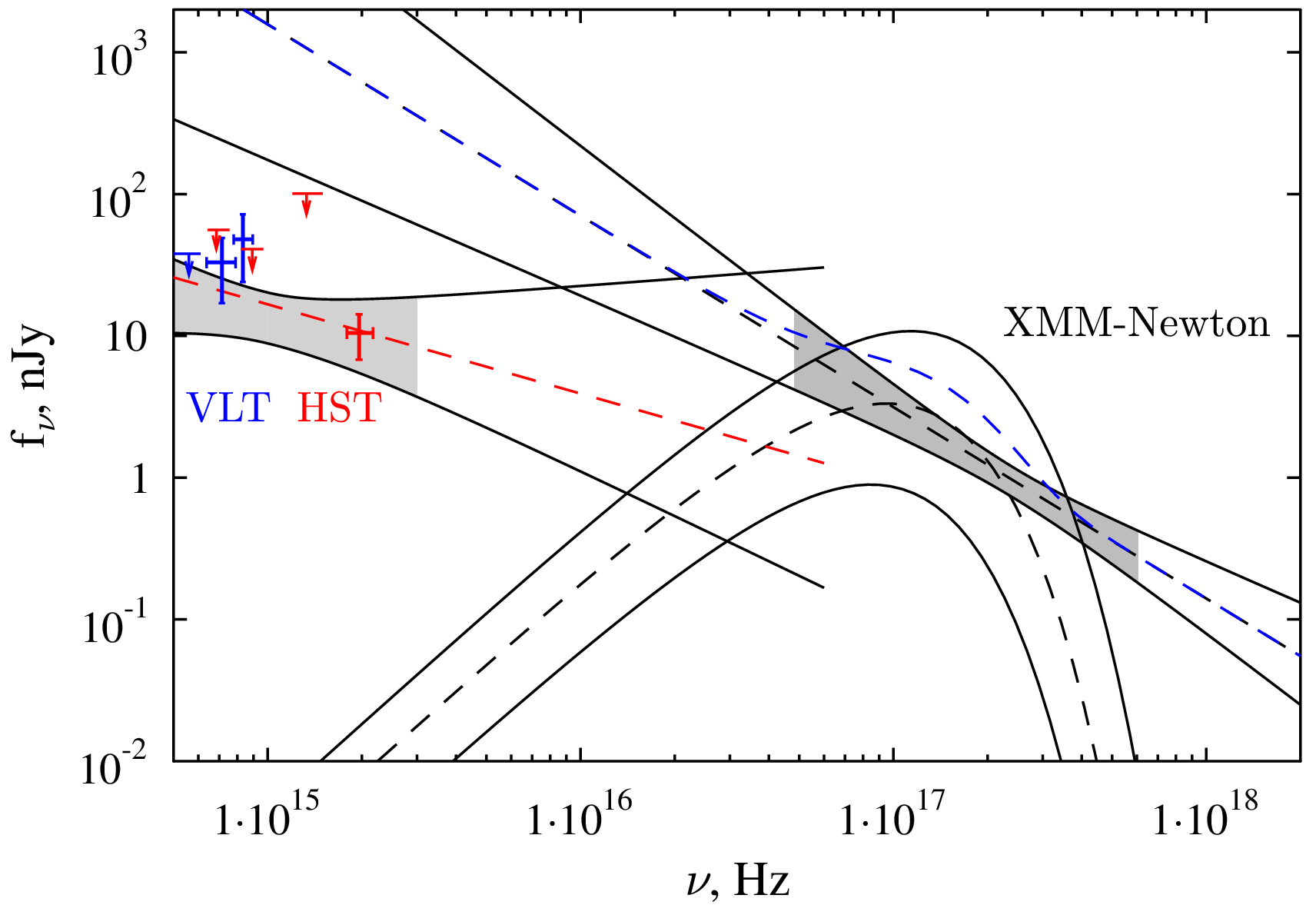}
\end{minipage}
\caption{
Absorbed (left panel) and unabsorbed (right panel) multiwavelength spectra of the pulsar. The {\sl XMM-Newton} X-ray spectrum is described by a combination of a 
PL component with a photon index $\Gamma=2.35^{+0.35}_{-0.40}$ plus a 
BB component with a temperature $kT=0.14^{+0.03}_{-0.02}$~keV, at a fixed hydrogen column density 
$N_{H}=1.4\times10^{20}$  cm$^{-2}$.
The dashed blue line in the left panel shows an extrapolation of the best-fit model spectrum to the optical-UV range, which strongly overshoots 
the observed optical-UV data points  shown in this plot.  
The red line in the right panel shows the best PL fit with $\alpha=-0.6$ to the optical-UV data points dereddened with $E(B-V)=0.02$, while black lines correspond to the 10\% and 90\% percentiles of the probability distribution 
of the optical-UV fit parameters. Blue dashed line shows the best-fit X-ray model spectrum, whereas black lines show its components and their uncertainties. 
\label{fig:8}}
\end{figure*}
Assuming both optical ($B$ and $U$) and FUV (F140LP) detections were real, we can fit the unabsorbed flux density points with a 
PL model: $f_\nu = f_0 (\nu/\nu_0)^\alpha$, where we choose the reference frequency $\nu_0 = 1\times 10^{15}$ Hz.
%
Following the approach suggested by   \cite{Sawicki2012} and 
developed by  \cite{Drouart2018}, 
in addition to the three detected SED data points,
we also 
included in the fits the four non-detections ($1\sigma$ upper bounds) 
in other filters presented in Table~\ref{tab:extinction}.  
Specifically, we used the  Python package {\tt Mr-Moose}  \citep{Drouart2018}, which allows data fitting in a Bayesian 
framework implementing the Markov chain Monte Carlo (MCMC) approach. 
To do that, we included FWHM of the {\sl HST} and VLT filters into the {\tt filters} directory of the {\tt Mr-Moose} distributive. 
For the MCMC convergence, we 
utilized 
1000 walkers and 1000 steps.  
The fitting parameters were the spectral index $\alpha$ and the PL normalization $f_{0}$, which were
allowed to vary in wide ranges between $-1.7$  and 1.1 for $\alpha$ and between $-31.2$ and $-30.5$ 
    for  $\log f_0$ (where $f_0$ in units of erg\,cm$^{-2}$\,s$^{-1}$\,Hz$^{-1}$). 
Their best-fit values with uncertainties, corresponding to 10 and 90 percentiles of the parameter distribution obtained by cumulative integration of the posterior probability density function,
are presented in Table~\ref{tab:extinction}.
Large uncertainties in the data 
result in large uncertainties of the PL parameters, and the parameter ranges  are almost insensitive to $E(B-V)$.
As an example,  we show the fit results for $E(B-V)=0.02$ in Figure~\ref{fig:7}, where 
the right panel 
shows 1D and 2D marginalized posterior probability distributions of the parameters reflecting  the fit convergence and quality. 
The fit results 
practically do not depend 
on the {\sl HST}/UVIS flux upper bounds, while they are  critically affected by the VLT upper bound in the $V$ band.   
We also tried to include into the fit  the VLT $U$ and $B$ 
band upper bounds 
of 2009, which  are consistent with the detections in these bands of 2000 (see Sec.\ 1).   This led  to a marginal flattening of the best-fit PL. For instance, we obtained $\alpha \approx -0.57$ instead of $-0.63$ without these upper bounds, for $E(B-V)=0.02$. Such a small difference is insignificant accounting for the large error budget (see Table~\ref{tab:extinction}).

 The obtained constraints on the spectral index of the pulsar's nonthermal emission  are compatible with (broader than) 
 a typical range
 $-0.7\lesssim \alpha\lesssim 0.2$
 for other pulsars observed in the optical-UV range \citep[e.g.,][]{2011mignani,2019mignani}.  
The best-fit PL parameters correspond to the optical-UV flux 
 $F({\rm 1500-6000\,\AA}) = 2.3\times 10^{-16}$ erg cm$^{-2}$ s$^{-1}$ and luminosity 
 $L({\rm 1500-6000\,\AA}) = 1.2\times 10^{27} d_{\rm 210\,pc}^2$ erg s$^{-1}$. Comparing the latter value with  the FUV luminosities 
 of pulsars 
 that have been detected 
 in this range  \citep{2019mignani},
 we see that J0108--1431 might be  the least luminous FUV pulsar.  
 On the other hand, the efficiency of 
 conversion of pulsar rotation energy to optical-UV radiation is $\eta_{\rm opt-UV} = L({\rm 1500-6000\,\AA})/\dot{E} = 2.4\times 10^{-4} d_{\rm 210\,pc}^2$. 
 Accounting for the distance and fit uncertainties,
 the efficiency could be anywhere in the range
 $0.7\lesssim \eta_{\rm opt-UV}/10^{-4} \lesssim 5.9$.
 Even for lower values from this range, 
 PSR J0108--1431 is the most efficient nonthermal   emitter  
 among 
 pulsars 
  detected in the optical-UV. 
  For instance, based on the data obtained by \citet{Pavlov2017}, the 17 Myr old  PSR B0950$+$08 shows  about two orders of magnitude 
  lower $\eta_{\rm opt-UV}$  of about  
  $6\times 10^{-6}$ in the same range.
  A typical optical efficiency range of pulsars is $10^{-7}$--$10^{-5}$, only the very young and much more 
  energetic Crab and B0540--69 pulsars have  efficiencies  comparable  to the 
  above estimate for J0108--1431 \citep{2006zhar,2010mpk,2015kir}.
  
  On the other hand, the pulsar is also a highly efficient nonthermal emitter in X-rays with 
  $\eta_{\rm X} \sim$ 0.003--0.006  \citep{Posselt2012}. 
  This results in the ratio of the nonthermal optical-UV to the X-ray luminosity 
  $L({\rm 1500-6000\,\AA})/L_{X} \sim 0.02$--0.04,
  marginally compatible with a typical range of 0.001--0.01 
  for pulsars observed 
  in the optical-UV and X-rays \citep{2004zp,2006zhar}. 


  It is interesting to compare the optical-UV spectrum of the pulsar candidate with the 
  X-ray 
  spectrum of PSR J0108--1431 obtained with \textit{XMM-Newton} \citep{Posselt2012, Arumugasamy2019}. 
  The latter presumably consists of a magnetospheric PL component 
  and a thermal component emitted from 
  hot polar caps. 
  Using the XSPEC tool (ver.\ 12.11.0)\footnote{See \url{https://heasarc.gsfc.nasa.gov/docs/xanadu/xspec}}, we fit
  the time-integrated spectra obtained  by 
  {\sl XMM-Newton} EPIC-pn and MOS1 instruments in the 0.2--10 keV range  with the absorbed PL+BB model at fixed $N_{\rm H}=1.4\times 10^{20}$ cm$^{-2}$, corresponding to $E(B-V)=0.02$.
  We obtained the photon index $\Gamma = 2.35^{+0.35}_{-0.40}$ (i.e., $\alpha_X = -\Gamma +1 = -1.35^{+0.35}_{-0.40} $), the PL normalization 
  $(1.4
  \pm 0.4)\times 10^{-6}$ ph cm$^{-2}$ s$^{-1}$ keV$^{-1}$ at $E=1$ keV, 
  the temperature $kT_{\rm BB} = 0.14^{+0.03}_{-0.02}$
  keV 
  ($T_{\rm BB}=1.6^{+0.3}_{-0.2}$ MK),
  and 
  the BB radius 
  $R_{\rm BB}= 22^{+23}_{-18} d_{210}$~m ($\chi_\nu^2 \approx 0.9$ for $\nu=20$ d.o.f). 
  The fit parameters are consistent, within the uncertainties (quoted above at a $1\sigma$ confidence level),
  with those 
  obtained by \citet{Posselt2012}, who 
  fixed photon index (at $\Gamma=2.0$) instead of $N_H$. 
  The obtained temperature is 
  within the range of typical BB temperatures of hot polar caps of old pulsars, while the BB radius is 
  a factor of 10 smaller than the ``canonical'' cap radius $R_{\rm pc} = R_{\rm NS} (2 \pi R_{\rm NS}/cP)^{1/2} \approx 238$~m for $P=0.808$~s, 
  assuming a 
  plausible intrinsic NS radius 
  $R_{\rm NS}=13$~km.  
  A similar discrepancy between $R_{\rm BB}$ and $R_{\rm pc}$ is observed in other old pulsars \citep[see][and references therein]{2020geppert,Posselt2012}. 
  %
  Figure \ref{fig:8}  
  shows the fit, for the absorbed and unabsorbed spectra, together with the PL fit to the optical-UV spectrum shown in Figure \ref{fig:7}. We see that the continuation of the X-ray model spectrum dominated by the PL component into the optical-UV lies well above the optical-UV detections and upper limits. 
  If we associate the optical-UV and X-ray PL components with the pulsar's magnetosphere emission, we can conclude that its spectrum steepens with increasing photon energy towards X-rays and  
  has a spectral break somewhere between the UV and soft X-rays,
  similar to other (younger) pulsars that were detected in both X-rays and the optical \citep[e.g.,][]{Kargaltsev2007}. 
  However,  better quality data in both ranges are needed to make a convincing multi-wavelength spectral analysis.  
  
  Based on the high optical through X-ray efficiency of J0108--1431, one can speculate 
  that for old pulsars the highest efficiency range 
  migrates from $\gamma$-rays towards lower photon energies.


  \subsubsection{Possibility of thermal emission in FUV and limits on surface temperature}
  
  The main goal of these observations was to constrain the NS surface temperature.
  First of all, we note that if the optical emission allegedly detected with the VLT  were thermal, then
  the FUV flux should 
  be much higher than either its presumably measured value or the upper bound, at any reasonable temperature and size of the emitting NS surface. 
 Therefore, we can rule out the temperature estimate by \citet{Mignani2008} derived from the assumption that the optical spectrum is a Rayleigh-Jeans part of thermal emission from the entire NS surface.


On the other hand, it is possible that the FUV emission is due, at least partly, to a thermal component. If the $U$ and $B$ detections 
are associated with the pulsar, 
then the relatively low upper limit on the $V$ flux does not allow a steep negative slope of the PL component, which leaves little room for the thermal component in F140LP.
If, however, the $U$ and $B$ detections are not associated with the pulsar
while the F140LP detection is,
then the F140LP flux  could be entirely thermal. 
As seen from Figure~\ref{fig:8}, it cannot come from  hot polar caps of the pulsar seen in X-rays, but can only come from a cooler bulk surface of the NS.
Assuming that the
spectrum of the NS surface emission 
is described by the Planck function,
$B_\nu(T) = (2h\nu^3/c^2)[\exp(h\nu/kT)-1]^{-1}$, 
its brightness temperature can be estimated 
from the observed flux density, 
\begin{equation}
f_\nu = (R_\infty/d)^2 \pi B_\nu(T_\infty) 10^{-0.4 A_\nu}\,,
\label{eq1}
\end{equation}
where 
$T_\infty=T/(1+z)$~K  and  $R_\infty=R(1+z)$~km  are the NS 
temperature and radius
as measured by a distant observer, $z=[1-2.953 
(M/M_\odot)(1\,{\rm km}/R)]^{-1/2} - 1$
is the gravitational redshift.
For the F140LP filter ($\nu_{\rm piv} = 1.96\times 10^{15}$ Hz, $A_\nu = 8.15  E(B-V)$), 
we have
\begin{equation}
T_\infty = \frac{9.42\times 10^4\,{\rm K}}{\ln\left[1 + \frac{187\,{\rm nJy}}{f_{\rm F140LP}}\left(\frac{R_{15}}{d_{210}}\right)^2 10^{-3.26 E(B-V)}\right]}\,,
\end{equation}
where  
$R_{15}=R_\infty/15\,{\rm km}$, and $d_{210}= d/210\,{\rm pc}$.

\begin{figure*}[t]
\begin{minipage}[h]{0.5\linewidth}
\includegraphics[scale=0.2,angle=0]{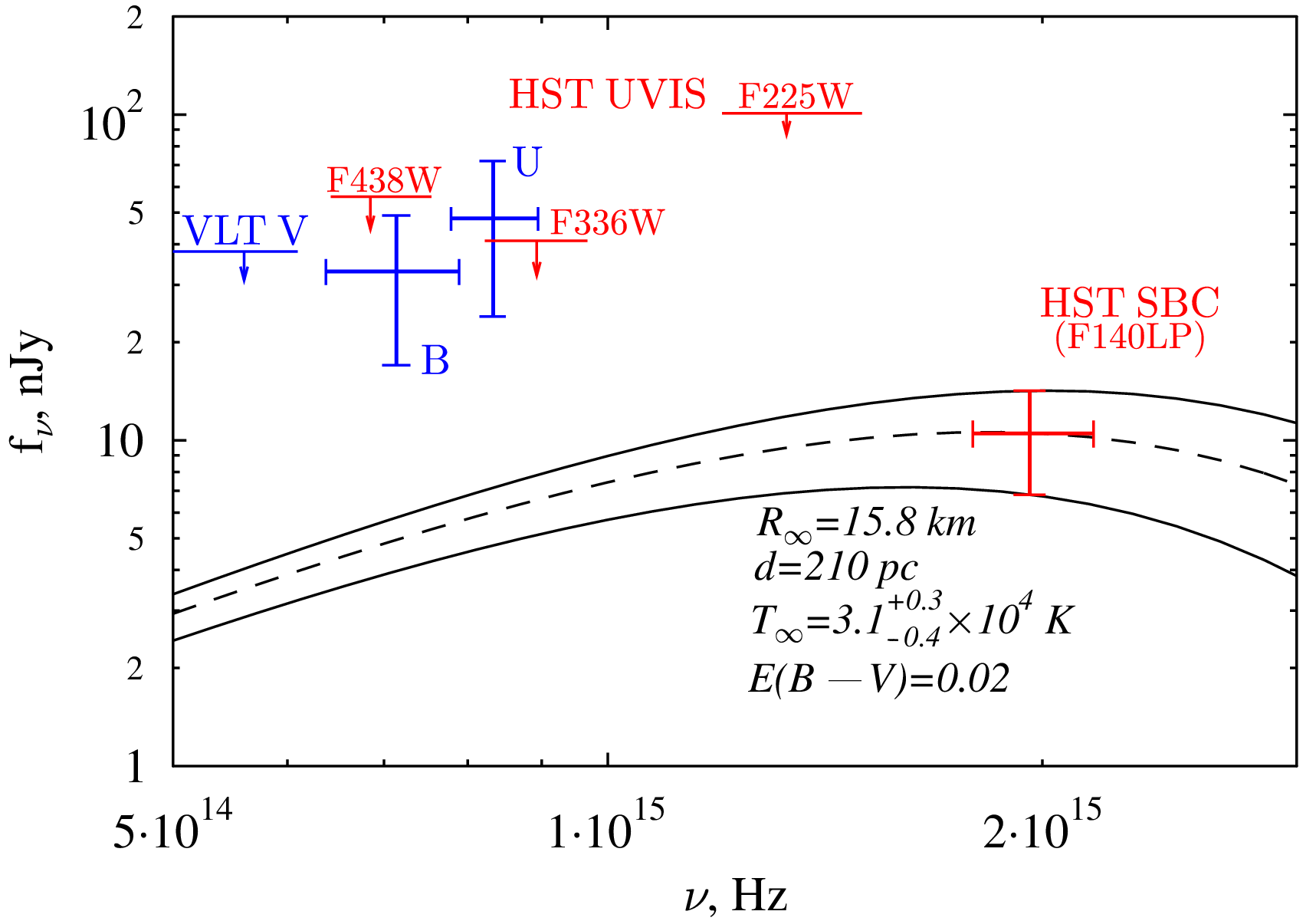}
\end{minipage}
\hfill
\begin{minipage}[h]{0.5\linewidth}
\includegraphics[scale=0.2,angle=0]{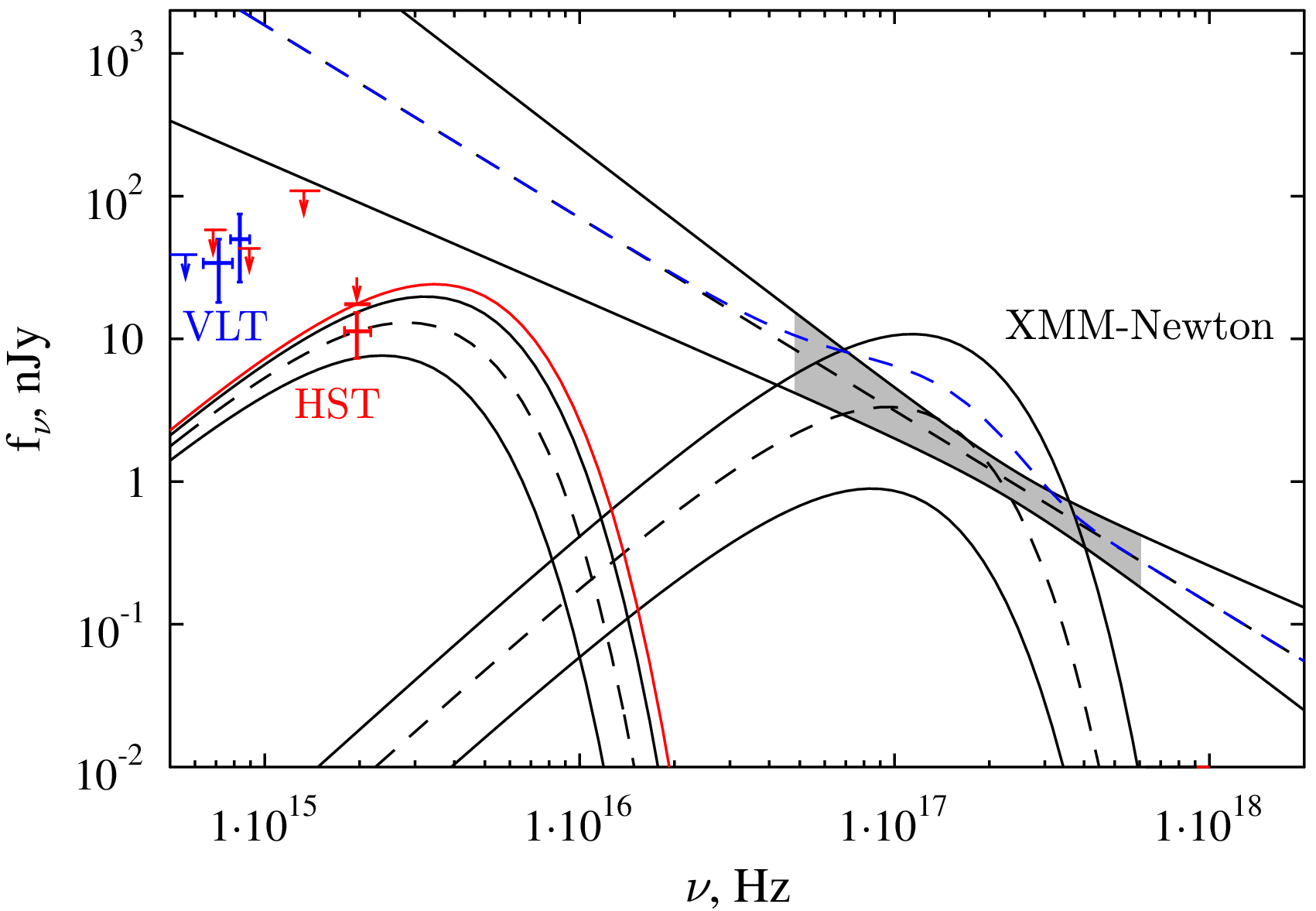}
\end{minipage}
\caption{
Limits on 
thermal emission from the NS surface.
{\sl Left:} 
Unabsorbed optical-UV blackbody spectrum assuming the F140LP flux is fully thermal, for
$R_{\infty}=15.8$ km, 
 $d=210$ pc, 
$E(B-V)=0.02$, and surface temperature $T_\infty=3.1^{+0.3}_{-0.4}\times10^4$ K (the temperature uncertainties 
correspond to the flux uncertainties). 
{\sl Right:} Dereddened optical-UV thermal spectra extrapolated towards higher energies, and the PL+BB fit to the {\sl XMM-Newton} data (the same as shown in Figure \ref{fig:8}). The blackbody spectrum from the entire NS surface shown by  black lines 
is plotted for $d=300$ pc, $R_\infty = 13.1$ km, 
$E(B-V)=0.03$,  $T_\infty = 4.8^{+0.7}_{-0.8}\times 10^4$ K. The red line corresponds to
a conservative upper limit of the NS surface temperature,
$T_\infty < 5.9\times 10^4$ K, when we consider the F140LP data point as an upper limit, for the same $d$, $R_\infty$ and $E(B-V)$.
\label{fig:9}}
\end{figure*}

For plausible $d=210$ pc, $E(B-V)=0.02$, and $R_\infty = 15.8$ km (which correspond to $R=13$ km at $M=1.4 M_\odot$), the measured FUV flux density
$f_{\rm F140LP} = 9.0\pm 3.2$ nJy 
yields $T_\infty = 3.1^{+0.3}_{-0.4} \times 10^4$  K ($T=3.8^{+0.3}_{-0.5}\times10^4$ K). 
The 
optical-UV part of this thermal spectrum with its uncertainties   
is shown 
in the  left panel of Figure \ref{fig:9}.
In the  right panel of this figure, we show the dereddened thermal spectrum at an upper end of plausible distances, $d=300$ pc, and 
a lower end of plausible radii,  $R_\infty = 13.1$ km  (corresponding to $R=10$ km at $M=1.4 M_\odot$ \citep{LattPrak2016} for the 
upper bound of the color index, 
$E(B-V)=0.03$; these parameters, correspond to a higher temperature, $T_\infty = 4.8^{+0.7}_{-0.8} \times 10^4$ K ($T=6.3^{+0.8}_{-1.1} \times 10^4$ K).
The spectrum is extrapolated   
towards 
X-rays, where the unabsorbed PL+BB fit of the {\sl XMM-Newton} spectrum of the pulsar   
is also shown. 
We see that emission from the NS surface 
with such a temperature would not be detectable in  the optical and X-rays.
If we assume that the F140LP detection was not real, then
the $3\sigma$ upper bound on the FUV flux density, $f_{\rm F140LP} < 14$ nJy, gives us an upper limit on the
temperature -- e.g., $T_\infty < 5.9\times 10^4$ K ($T<7.7\times 10^4$ K) at $d=300$ pc, $R_\infty = 13.1$ km, $E(B-V)=0.03$ (the corresponding thermal spectrum is shown by the red line in the right panel of Figure \ref{fig:9}).

It is interesting to compare the obtained constraints on the surface temperature of the 
 196
Myr old PSR J0108--1431 with those for other old pulsars. Based on the upper limit, $T_\infty \lesssim 6\times 10^4$ K, we can conclude that PSR J0108--1431 is colder than the 17 
Myr old PSR B0950+08, the only old ordinary pulsar whose thermal emission has been detected, with $T_\infty$ in the range of
(1--$3)\times 10^5$ K \citep{Pavlov2017}. 

If the F140LP detection of thermal emission from PSR J0108--1431 was real, then 
the plausible temperature range, 
$T_\infty \approx (2.7$--$5.5)\times 10^4$ K, is just slightly above the conservative 
upper limit, $T_\infty \lesssim 3.2\times 10^4$ K, for the 330 Myr old PSR J2144--3933 \citep{Guillot2019}\footnote{\citet{Guillot2019} present the upper limit $T\approx 4.2\times 10^4$ K 
for 
unredshifted temperature assuming an NS with $R=10$~km and $M=1.4 M_\odot$.}. 
The highest temperature of this range is lower than the surface temperature, 
$T_\infty \approx (1.2-3.5)\times 10^5$~K,  
of the 7 Gyr old  nearest millisecond (recycled) pulsar 
(MSP) J0437--4715 \citep{2004karg,2012durant,2019gonzales}.
It means that either thermal evolution of NSs is not monotonous or it proceeded differently for J0108--1431 and J0437--4715, 
e.g., because  old ordinary and recycled pulsars 
have some different properties, including periods and their derivatives, magnetic field strengths and masses \citep{2015gonz}. 

Confirmation of the possible detection of thermal emission from J0108--1431 would mean that it is the coldest NS whose thermal emission has been detected, but 
its temperature is still high enough to support the idea that NSs do not just cool passively 
but some heating mechanisms strongly affect their thermal evolution. 
According to passive cooling scenarios, cooling of isolated NSs 
becomes exponentially fast
at ages of  a few  Myr after which  thermal emission from their surfaces 
becomes undetectable \citep[e.g.,][]{2004yp}. 
However, this rapid cooling can be partly  compensated by a number of heating mechanisms.
One of them is the so-called rotochemical heating due to composition changes (such as the neutron beta decay) forced by density increase as the centrifugal force decreases in the course of NS spindown 
\citep{
1995reiseng,2005fr}.
This heating mechanism has a minor effect on surface temperatures of young NSs but its contribution can  dominate at ages $\gtrsim 10$ Myr. 
Another important mechanism is ``frictional heating'' caused by interaction of vortex lines of the faster rotating neutron superfluid with the slower rotating normal matter in the inner NS crust \citep{1984Alpar,1999ll}.


According to the top panel of  Figure 5 of \citet{Guillot2019}, the
upper bound on the (unredshifted) surface temperature of PSR J0108--1431, $T < 8\times 10^4$ K,
is consistent with 
the values predicted by the models of rotochemical heating by \citet{2010GR} for a 200 Myr old pulsar with the surface magnetic field of $2.4\times 10^{11}$ G and initial period at birth of 1 ms, assuming either modified Urca reactions or direct Urca reactions with additional frictional heating with excess angular momentum $J=1\times 10^{44}$ erg s (the predicted temperatures are very close to each other at these parameters). 
This is similar to the younger PSR B0950+08. 
However, if the FUV thermal emission was actually detected, then the observed temperature range $T\sim (3$--$7)\times 10^4$ K 
lies between  these predictions and the low temperature  boundary provided by the direct Urca models 
without frictional heating. This would mean that  frictional heating is less 
efficient in PSR J0108--1431 than in  PSR B0950+08.
To obtain tighter constraints on heating mechanisms, one should re-observe PSR J0108--1431 with deeper exposures as well as observe more old pulsars in the optical-UV.




\section{Conclusions}

We observed the field of the nearby  
196 Myr 
old PSR J0108$-$1431 with the {\sl HST} in four optical-UV bands. 
We detected a point-like FUV source in the 
F140LP band at about 3$\sigma$ significance
level with coordinates coinciding with 
the position of the pulsar within the 1$\sigma$ uncertainty
of 0\farcs2. 
We consider this source as a possible FUV counterpart of 
PSR J0108$-$1431. Also, we placed upper limits on 
the flux densities of the pulsar in the F225W, 
F336W and F438W bands. 
Using more accurate astrometry, we confirmed the 
$3\sigma$ detection of the optical source at the pulsar position in year 2000 in the VLT {\bf $U+B$} filters, 
 and its upper limit in the $V$ filter,
reported 
earlier by \citet{Mignani2008}.    

Assuming that the possibly detected F140LP, $U$ and $B$ emission comes from the pulsar counterpart,
we analyzed its 
multi-wavelength 
spectral energy distribution,
 including the 
archival X-ray data obtained with 
\textit{XMM-Newton}. We found that
the spectral flux density distribution can be described by a 
power-law model in the optical-UV part, suggesting its magnetospheric origin.
The spectrum becomes steeper in X-rays, implying 
a spectral break between the UV and X-ray ranges. 
Such 
behavior is typical for pulsars observed in both ranges. 
The pulsar has a record high efficiency, $\eta\sim 10^{-2}$, of 
transformation of the 
spindown power to nonthermal (magnetosperic) radiation 
in the optical-UV through X-rays.

In the FUV band, the pulsar emission 
might be dominated by thermal emission 
from the bulk of the NS surface.
If this is the case,
the NS surface temperature is
in the range of (3--6)$\times 10^4$~K,
as seen by a distant observer 
-- the lowest NS temperature ever measured.
At the same time, it is much higher 
than predicted by scenarios of passive NS cooling at this NS age,
which could be due to heating mechanisms operating in the NS interiors. A conservative consideration of the FUV data point as  an  upper bound yields the 3$\sigma$   upper limit on the NS  brightness temperature,   $T_\infty < 6\times 10^4$~K. 


Detection of 
PSR J0108--1431 in the optical and 
UV bands at higher significance levels is needed to confirm the counterpart and study its properties.

\acknowledgments
 We thank the referee 
for useful comments. 
Support for \textit {HST}
program \#14249 was provided by NASA through a grant from
the Space Telescope Science Institute, which is operated by the
Association of Universities for Research in Astronomy, Inc.,
under NASA contract NAS 5-26555.
We thank Bettina Posselt for providing the reduced {\sl XMM-Newton} data. RPM is grateful to Denise Taylor (STScI) for support during the \textit{HST}  observations.

\facilities{\textit{HST}/WFC3; \textit{HST}/ACS; VLT/FORS; \textit{XMM-Newton}; \textit{Gaia}}

\software{This research made use of the following softwares and packages: IRAF \citep{1986Tody, Tody1993}, XSPEC \citep{1996Arnaud}, EsoRex \citep{2015ESO}, Mr-Moose  \citep{Drouart2018}}

\bibliographystyle{aasjournal}
\bibliography{rnaas}

\end{document}